\RequirePackage{amsmath}
\documentclass[runningheads]{llncs}

\usepackage[left=1.75in, right=1.75in]{geometry}
\usepackage{graphicx}
\usepackage{amssymb}
\usepackage{lineno}
\usepackage{multirow}
\usepackage{mathtools}
\usepackage{hyperref}
\usepackage{color}
\usepackage{tikz}
\usetikzlibrary{positioning, shapes.geometric, arrows.meta}
\usepackage{verbatim}
\usepackage[square,sort,comma,numbers]{natbib}
\usepackage{tabularx} 

\begin{document}

\title{Social rules for agent systems\thanks{This research was partially supported by the Wallenberg AI, Autonomous Systems
and Software Program (WASP) funded by the Knut and Alice Wallenberg Foundation.}}
\author{Ren\'e Mellema\inst{1}\orcidID{0000-0002-4138-937X} \and Maarten Jensen\inst{2}\orcidID{0000-0002-6117-8019} \and Frank Dignum\inst{3}\orcidID{0000-0002-5103-8127} }
\authorrunning{Mellema et al.}
\institute{Ume\aa\ University, Sweden\\ \email{rene.mellema@cs.umu.se} \and \email{maarten.jensen@cs.umu.se} \and \email{dignum@cs.umu.se}}
\maketitle

\begin{abstract}
    When creating (open) agent systems it has become common practice to use social concepts such as social practices, norms and conventions to model the way the interactions between the agents are regulated. However, in the literature most papers concentrate on only one of these aspects at the time. Therefore there is hardly any research on how these social concepts relate. It is also unclear whether something like a norm evolves from a social practice or convention or whether they are complete independent entities. In this paper we investigate some of the conceptual differences between these concepts. Whether they are fundamentally stemming from a single social object or should be seen as different types of objects altogether. And finally, when one should which type of concept in an implementation or a combination of them.
\end{abstract}

\section{Introduction}
In the last twenty years several social rules such as conventions, norms and recently also social practices, rituals and habits have been used to model agent behaviour in multi-agent systems. The main intuition behind the use of these rules is that agents are at least partially autonomous and thus their behaviour cannot just be constrained through some hard constraints. Moreover, the social rules have not only a constraining character, but also have a motivational component in that they are part of the deliberation process. E.g.\ a norm to keep to the traffic rules as a bicyclist can lead to a plan to leave home early for a meeting in order to have enough time to navigate the traffic in a legal manner. The fact that the social rules have different aspects and consequences makes them difficult to implement in a uniform way as that implementation depends on what kind of internal mechanisms are available in the agents as well as in the environment. Thus different applications have concentrated on different aspects of the social rules, while most applications only use one of the social rule types.

Although for many applications one can suffice with modelling the social rules as if they are static, it is clear that in reality they are not. Norms and conventions emerge and adapt to the environment. In \cite{castelfranch-giardini-lorini} it is argued that some expectations can grow to norms via conventions. Thus the social rules stem from a common root of expected behaviour and can grow into different forms dependent on their use and the environment. This is an interesting view point as it would shed light on commonalities and differences between the social rules. Are all social rules starting with common expectations of some behaviour and then evolving into more specific rule types by adding specific types of characteristics?

In this paper we discuss the different types of social rules. In particular we will look at social practices, conventions, social norms, moral norms, legal norms, rituals and habits. The reason why we constrain ourselves to only the social rules is that we see them as a class of social objects on their own. While it would have been possible to also include other social concepts, such as institutions and organizations, we see these concepts as of a different kind from the social rules. They have therefore been left out of this analysis, but we plan to incorporate them in the current framework in later work.

We will give a short overview of all the selected social rules in section 2.
In section 3 we will compare the different types of social rules and particularly discuss whether some types of rules evolve from other types, how this happens and what the consequences are of this evolution. The main contribution of the paper will be a better understanding of the unique characteristics of each social rule type and the implications of these characteristics for implementations. These implications will be discussed in section 4. Note that in this paper we will not use any formalism to describe the social rules. This is done on purpose as it would inadvertently already bias any discussion to those features of the social rules that are more prominent in the formalism. This does not mean that we believe that the social rules cannot be formally modelled. In~\cite{dignum2020} we give a summary of formal models of the social rules and why these formalisms are especially appropriate for each of them. We will give a short conclusion in section 5.

 \section{Types of Social Rules}
 Social rules have developed in all societies around the world. The general intuition is that these social rules will guide behaviour of people in an increasingly complex environment. Having social rules allows people to create expectations about the behaviour of other people and thus afford a more feasible planning of their own behaviour. This \emph{standardization} of behaviour is one of the most important functions of all the different types of social rules.

 \subsection{Social practices}
 The concept of a Social Practice is described and investigated in Social Practice Theory. Social practice theory comes forth from a variety of different disciplines. It started from philosophical sociology with proponents like Bourdieu~\cite{Bourdieu1976} and Giddens ~\cite{Giddens1979}. Later on Reckwitz, Shove and Schatzki~\cite{reckwitz_toward_2002,shove2012,Schatzki2012} have expanded on these ideas. An important claim is that human life should be understood in terms of organized constellations of interacting persons. Thus people are not just creating practices, but our deliberations are also based on the fact that most of our life is \emph{shaped} by social practices. Thus we use social practices to categorize situations and decide upon ways of behaviour based on social practices. This becomes immediately clear when we think about our everyday life. When we get up we have a routine for getting dressed and having breakfast with the family. This is a social practice that we perform every day, without thinking about it. Then we go to school or work, have dinner with the family and have some more activities (or work) afterwards until we all go to bed at about the same time every day. Thus most of our daily lives are governed by practices that are also linked together and that we follow without having to deliberate or plan them. Using social practices in this way simplifies an otherwise very complex life by providing a kind of standard interaction pattern including resources and context needed for those interactions.

 In our view, social practices always require some direct level of interaction between people, which is not always the case in existing social practice literature. For example, in~\cite{higginson_diagramming_2015}, they also call doing the laundry a social practice, even though this only involves one individual.

 According to~\cite{reckwitz_toward_2002,shove2012} a social practice consists of three parts:
 \begin{itemize}
     \item \textit{resources}. The resources refer to the time, place, and objects used in a social practice and the types of actors involved.
     \item \textit{activities}. The activities of the practice refer to the types of actions that are involved in the practice and possibly their ordering.
     \item \textit{meaning}. The meaning of the social practice indicates the social aspects of the practice. It indicates which social interpretations are linked to certain actions in that context and also social effects or purposes of the practice.
 \end{itemize}
 The \emph{resources} for a shopping spree down town would be the time available to go shopping, the shops that one wants to visit (clothes shops, warehouses, restaurants, hardware stores, etc.) and the employees in the shop.\\
 The \emph{activities} denote the kinds of activities that can be expected. They do not dictate a complete workflow or script for the social practice, but rather indicate the landmarks for synchronization and which actor (or role) in that practice has the initiative in each activity. The individuals participating in the social practice are free to act in any way that conforms to the practice, and even to violate it if they wish. So, going shopping with some people can involve window shopping, having a coffee in a caf\'e, and even actually buying particular products. These activities can be done in any particular order and not always all of them are performed.\\ The \emph{meaning} of a social practice couples the interaction perspective of it with the social perspective of the practice. Thus going shopping is not just about buying things. People can go ``shopping'' as a means to have a social meeting. It can be a way to find out what is the new fashion or the latest gadgets; but it can also be a way to align your tastes with those of your shopping companions.

 \subsection{Conventions}
 The prototypical example of a convention is that of walking on the right hand side of the pavement. There is no intrinsic value of walking on the right or the left side of the pavement. The main reason for such a convention is to walk all at the same side, which makes coordination in crowded areas much easier. In general (!) conventions have an externality effect. I.e.\ if more people conform to the convention it becomes more profitable to also conform to it. In the other hand, often the violation of a convention leads to an immediate negative effect. E.g.\ walking on the opposite side of the pavement will lead to collisions and thus will be avoided automatically. Conventions are justified through their immediate positive impact on the coordination of frequently occurring (inter)actions. Thus they can be seen as solving a coordination problem through the creation of common expectations about behaviour in that coordination. This is exactly how it is defined by Lewis in~\cite{Lewis1969}:\\
 A behavioural pattern C in a population P in a recurrent situation S is a \emph{convention} if it is \emph{common knowledge} in P that in situation S
 \begin{itemize}
     \item everyone conforms to pattern C
     \item everyone expects everyone else to conform to C
     \item everyone prefers to conform to C on condition that the others do so
 \end{itemize}
 For anyone who is familiar with game theory the above definition looks very similar to a Nash equilibrium of a game where the participants know the pay-off matrix.

 \subsection{Norms}
 Although norms have been studied for a long time in many different disciplines there is still little consensus about a definition or even a classification of norm types. We will not venture into that discussion but use a rather simplified classification of norm types that suits our purposes best. We distinguish social norms, moral norms, legal regulative norms and constitutive norms.

 Social norms are usually used in contexts where the change of norms is the focus of attention. Thus they are described in terms of interactions and how behavioural patterns emerge and depend on context and interaction mechanisms. In social science norms are used in a descriptive form, indicating the ``normal'' behaviour in some time period and what the consequences of following this norm are for the society. E.g.\ the norm in The Netherlands used to be that women do not have a paid job, but stay home and take care of the children. This has impacted the way society functioned for many years. The norm has now changed to one in which both men and women often work part-time. For social scientists the research questions are what might be the consequences for the children, or the consequences for work practices of this shift. In the corresponding research in social simulations the emphasis is on the behavioural pattern that is described by the norm and which parameters in the environment or the preferences of the individuals might cause the change of behavioural patterns. However, the norms are not explicitly modelled in the system but interpreted as being the normal behaviour that is observed.

 When social norms are established for some time they also might become moral norms. I.e.\ the behaviour that fulfils the social norm is seen not just as normal, but also as the \emph{right} behaviour. Thus the norm that women in The Netherlands did not have a paid job became a moral norm where women were supposed to be taking care of the children, because it was morally \emph{better}. Thus moral norms have an extra argument to follow them, since everyone is following them and it is \emph{good} to follow them.
 Moral norms concern moral judgements, and are founded in the values that a person has. Since they are not necessarily based on behaviour that a person sees around them, it is also possible for a person to have moral norms that are not generally held by society at large, or even by anyone else they know. Despite this, people do use moral norms to judge the behaviours of others, even if those others do not hold the same norm. This can, for example, be seen in the feminist protests last century which were geared to give women more rights and possibilities to work. Especially some fundamental religious groups opposed these rights. Here different groups clearly value different things, which has lead to different moral judgements on the subject.

 Usually, when norms get institutionalized to become laws or regulations they tend to become more abstract and the emphasis in their formulation shifts from regulating specific (inter) actions to desired situations. This shift allows (in general) to cover more situations with a norm, because a desired state can usually be achieved in many different ways. We now leave open how the state is achieved and thus can cover more situations. Again, this is a general heuristic, but it reinforces the underlying idea that norms are protecting the general interests and values of society (or at least part of society) and thus are meant to cover generalized and abstract situations rather than specialized individual contexts.
 The concept of a formal law emphasizes the sharedness of the norm, that is not easily changed, has precise consequences described on violation and is expected to be known by all. Explicit punishment and repair clauses are connected to these norms. It is also in this context that we encounter the term \emph{regimented} norms. These are norms that can be effectively controlled and thus enforced, such that violation is impossible. E.g.\ if it is forbidden to park in front of the school when dropping off the children, the school can regiment this norm by physically putting some barrier around the parking place, preventing a car to park there. It is also from this perspective that constitutive and regulative norms are distinguished. The constitutive norms define the terms and conditions of the regulative norms. E.g.\ an electric bike counts as a bike (rather than a self-propelled vehicle). Or, pushing the ``buy'' button on the screen of an webshop counts as accepting the terms and conditions of the sale. These definitions are also called norms, because they also influence our behaviour (indirectly) as they define the shared interpretation of actions and terms in a certain context. So, normally an electric bike is considered a bike under the Dutch traffic law. This definition can also change over time and context. E.g.\ one might not be obliged to wear a helmet on a bike, but is obliged to wear a helmet on an electric bike.

 \subsection{Rituals}
 We include rituals because in recent years it has been used especially in the context of virtual characters playing roles in culturally distinct worlds (see e.g.~\cite{Mascarenhas2009,degens2014}). In some sense rituals are close to social practices as they also describe interactions in which there are many expectations about the particular behaviour of each party involved. However, according to~\cite{Bell1997} rituals are:
 \begin{quote}
 Sequences of activities involving gestures, words, and objects, usually performed in a sequestered place, and performed according to a set sequence. Rituals may be prescribed by the traditions of a community, including a religious community. Rituals are characterized but not defined by formalism, traditionalism, invariance, rule-governance, (sacral) symbolism, and performance.
 \end{quote}
 So, the formalized nature and invariance of the actions performed during the ritual are emphasized plus the symbolic character of the ritual. This latter aspect is emphasized by Kyriakidis~\cite{kyriakidis2007} who states that rituals are used to distinguish in-group and out-group members.

 From these definitions and descriptions it becomes clear that rituals are not so much a means for efficient coordination of activities, but have a primarily social and symbolic function. Thus the particular actions are only important in as far as they signify (and can be recognized as doing so) the social purpose that is achieved with the ritual. In order for the social, symbolic nature of a ritual to be easily recognized, usually the actions performed are special or have clearly no direct functional purpose in the context in which they are being executed. E.g.\ a special handshake symbolizes that the ones using it are members of a group. They are the only ones that know exactly what to do and the handshake is also not normally used by others. Usually the participants in the ritual believe that the ritual is in fact the only way to establish the social purpose they achieve. Thus the ritual becomes equated with the social state it achieves and is a kind of manifestation or symbol of it. E.g.\ initiation rituals of student associations, but also initiation rituals when becoming an adult, wedding ceremonies, burial rituals, etc.

 \subsection{Habits}
 Whereas rituals can be seen as social practices that focus exclusively on the social aspect of the practice, habits can be seen as social practices that focus almost exclusively on the individual cognitive and action aspect of the practice. According to~\cite{andrews1903} habits are a more or less fixed way of thinking, willing or feeling acquired through previous repetition of a mental experience. This definition from 1903 emphasizes the mentally set way of deliberation concerning a habit. Although habits are a very individual pattern of behaviour they still can be used in social interactions as other people can expect someone to stick to his habits in a certain context and thus anticipate his behaviour. One could even argue that from the outside social practices look like the synchronous performance of habits of the people involved in the practice. However, although this might look so from the outside (looking at the interaction), the participants of a social practice are aware that they perform their actions as part of the social practice and also monitor the behaviour of the other participants to check if they comply to the expectations and whether the purpose of the social practice is achieved. This awareness of the social context and its consequences are not part of habits, which are mainly unconscious.
 Some aspects that get special attention in the literature on habits are habit formation and habit breaking. In~\cite{lally2010} some empirical work is discussed that indicates how long it takes persons to form a habit. Although there is considerable individual variance, many people form a habit within a couple of weeks if they perform an activity every day. In~\cite{Wood2007} it is argued that when a habit is formed the connection between the original goal for which the behaviour was performed and the plan are slowly disconnected. It is argued that this disconnection is one of the main reasons why it is difficult to subsequently change a behaviour. In~\cite{mercuur_changing_2017} we show that once such a disconnection is made, just getting information about negative effects of the habit does not lead to behavioural changes. This shows that the behaviour no longer is the consequence of a conscious deliberation about a course of action for a specific goal, but rather an automatic behaviour that is not influenced by any current beliefs.
 Most of the literature on habits focuses exactly on how (bad) habits can be changed. Instead of just trying to stop a bad habit it is better to replace a bad habit by a good one that is triggered by the same cue as the bad habit. Thus rather than going for a snack in the middle of the afternoon one could have some fruit available to eat at that time.
 Because habits are purely individual, deviations from habits are not really important from the social perspective. A deviation or obstruction of a habit has as only consequence that an undeliberated behaviour cannot be performed and deliberation has to take place about the formation of a plan to achieve the goal or react to the current situation.

 \section{Comparison of Social Rules}
 Now that we have given an overview of many social rules we will discuss possible relations between them in this section. We will look at the similarities, differences and whether one rule could emerge from another rule. Answering these questions will create a foundation for a combined implementation of these social rules. One thing to note is that, while we will discuss the differences between the social rules, we do not see the types as mutually exclusive, since we believe that the changes described happen gradually. Furthermore, we also believe that a certain behaviour can be governed by different social rules at the same time.

 If we take the example of walking in a pedestrian area we can see the social practice arises to keep to the right of the street, in countries where people drive on the right. Most people keep right just because they are used to keep right. However, when almost everyone keeps right when walking in this area it also becomes easier to walk when you keep right. Thus, a convention might arise due to the externality that keeping the same side as everyone else makes moving easier, since it leads to less bumping into each other. If the convention exists for some time it can also turn into a social norm, meaning it becomes good to keep to the right. Not just because everyone does it, but because keeping right is supposed to be better than swerving around or keeping left. If keeping right became a social norm we might observe people walking the street in the middle of the night, while no one else is there, still keeping to the right of the street.

 \subsection{Graphical relation of the social rules}
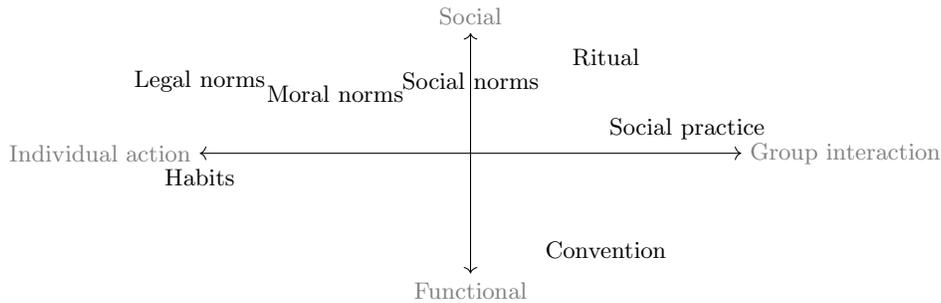
\begin{figure}[!ht]
     \centering
     \begin{tikzpicture}[scale=2, axis label/.style={gray}]
             \newcommand\xlen{1.8}
             \newcommand\ylen{0.8}
             \draw[<->] (0, -\ylen) -- (0, \ylen);
             \draw[<->] (-\xlen, 0) -- (\xlen, 0);
             \node[axis label, below] at (0, -\ylen) {Functional};
             \node[axis label, above] at (0, \ylen) {Social};
             \node[axis label, left] at (-\xlen, 0) {Individual action};
             \node[axis label, right] at (\xlen, 0) {Group interaction};
             \node at (-\xlen, -0.2*\ylen) {Habits};
             \node at (0.5*\xlen, 0.8*\ylen) {Ritual};
             \node at (0.8*\xlen, 0.2*\ylen) {Social practice};
             \node at (0.5*\xlen, -0.8*\ylen) {Convention};
             \node at (0, 0.6*\ylen) {Social norms};
             \node at (-0.5*\xlen, 0.5*\ylen) {Moral norms};
             \node at (-\xlen, 0.6*\ylen) {Legal norms};
     \end{tikzpicture}
     \caption{Representation of relation between the discussed social rules.}
     \label{fig:combined_social_rules}
 \end{figure}

 In figure~\ref{fig:combined_social_rules} we have represented the focus of the various social rules we use in this paper as a coordinate system with two axes. The horizontal axis represents how many people need to be involved when the rule is put in practice, and goes from individual action on the left to group interaction on the right. The figure shows that legal norms do not depend on many people following it. The decision to follow a legal norm is an individual one. It is exactly the fact that a legal norm is institutionalized and formalized that withdraws it from individuals following it or not. In the other extreme are social practices that only exist by virtue of groups of individuals practicing them. If people do not practice them, they disappear. This is opposed to legal norms that are not dependend on this feature.\\
 The vertical axis contrasts the functionality of a rule with its social effect. With functionality we mean that actions have a direct physical effect, such as two people bumping into each other, whereas social effect refers to for example the relations between people and their opinions of each other. This means that conventions, which have a direct physical effect but no social gain are placed at the bottom of the picture, whereas rituals, where the functional effect is insignificant, get placed at the top.

 \subsection{Social norms, legal norms, and moral norms}
 Since social norms, legal norms, and moral norms are all norms, they share a lot of similarities that set them apart from the other social rules. They all describe what an agent ought (not) to do, regulating the agents behaviour. All of them can be violated, after which a (social) sanction follows for the violating agents. They are all abstract to a certain level, specifying what kinds of behaviour should be done, or which situations are desirable. Therefore, we will mostly look at their differences. For this we will expand upon the distinctions already made in~\cite{Brennan:2013gi}.

 We will start with the distinction between social norms versus moral and legal norms. The difference between these two groups is that social norms are bound to the perceived existence of a practice, whereas the other two need not be. This means that the agents in the society have to believe that there is some social practice/convention/etcetera that they are obliged to follow because they believe other agents do so too. In the case of moral and legal norms the behaviour of other agents is less relevant, and may not have an influence at all.

 When we contrast legal norms versus social and moral norms, the biggest distinction is that legal norms are enforced by an institutionalized authority, but moral and social norms are enforced by the society directly. Normally this authority is instantiated and granted rights by the society itself to represent and protect its values and interests. Because of the level of indirection involved in this, the sanctions given by the authority are different from those given by the community itself. Sanctions from the community mostly lower an individual's social standing, whereas sanctions from the authority tend to be more directly harmful for the agent, such as fines or social service.

 The difference between moral norms versus social and legal norms, which was not mentioned in~\cite{Brennan:2013gi}, is that moral norms are more personal than social and legal norms. Both social and legal norms specify something about and for the community, what happens in and what is deemed best by the authority in power respectively. Moral norms, on the other hand, are directly tied to an agent's value system, and specify what that agent thinks is morally good. While most agents within a society will have similar value systems and a lot of shared values~\cite{Shalom2001cultures}, not all of them will overlap. An agent can have personal values that deviate from the shared values of the group it belongs to. E.g. wearing a helmet on a bike might be seen by society as morally good as it reduces the risk of injuries when falling and thus is part of a persons responsibility towards society to behave carefully. However, Dutch people living abroad might have a personal moral norm that states that wearing a helmet on a bike is unnecessarily cautious, because they never get injured when falling and wearing a helmet restricts their contact with the environment reducing their biking skills.

 Because the different norm types also share so many similarities, it is also possible for a norm to change from one type to another. We will start with the easiest direction to discuss, from a moral or social norm to a legal norm. This process tends to be quite explicit, since legal norms are normally written down and need to be properly communicated. This can happen when a society's authority has determined that some social or moral norm is very important or needs active enforcement from the authority. During this process, the norm normally becomes a bit more abstract so it covers more situations.

 Social norms can also transform into moral norms. This can happen when the social norm has been seen as inherently good by a society, while the practice that supports it has fallen out of fashion. In these cases the norm changes from ``good because many people do it'' to ``good because it is inherently good''. The norm then normally also starts to describe a situation that should be achieved, leaving the exact execution of the actions to get there up to the agents themselves.

 Legal norms can also become social or moral norms in cases where the legal norm in question is abolished by the authority, but the norm lives on ``by habit'' in that society. An example in the Netherlands is the raised speed limit on some highways, which went from 120 km/hr to 130 km/hr. Especially when the limit had just been heightened, a lot of people still adhered to the old limit of 120 km/hr.

 The only direction that is not discussed yet is from moral to social norms. While in theory this could happen, when a moral norm has a single behaviour associated with it and is no longer seen as intrinsically good, we do not expect that this occurs often. Since  moral norms tend to be more lenient in the behaviours they allow it is unlikely that there would be only one practice within a community that is associated with it, and if the norm is no longer seen as intrinsically good, then there is little reasons the agents have to keep up the behaviour, except for it being habitual.

 \subsection{Social practices, conventions, and norms}
 From social practices, conventions, and norms, the social rule with the most elements is the social practice. Because of this, one could say that they can be seen as the basis of all social rule types. Besides the interaction (or activity) specification they also contain the resources (including place and time) that are used in the interaction and there is an explicit meaning of the practice. These elements make the practice more situated and bound to specific situations. Where conventions and social norms assume the rule to be widely followed in society, a social practice can in principle be limited to a repeated interaction between two people. The fact that the meaning is explicit part of the practice indicates that it is not inherent to the interaction itself, but rather comes from the (social) interpretation of the interaction. E.g.\ suppose some people start driving 100km/hr on a 120 road. They do this because it is better for the environment.  At a certain moment the practice of driving a 100km/hr can have as meaning that you care about the environment. It can be that people drive 100km/hr just because they are distracted or for some other reason. However, looking at the practice from outside we cannot observe these motivations. The meaning is inferred from the practice itself.

 We do not have a similar element for conventions and norms. Conventions do not carry any meaning. They are just a convenient equilibrium that defines a certain type of behaviour is optimal in an environment given that others behave in a certain way. Thus walking on the right has no particular meaning. It is just the most convenient way to walk given that others also keep to the right.\\
 Social norms also do not carry this type of meaning. For norms the meaning of the norm comes from the fact that everyone follows the norm and one wants to conform to what other people do. So, one could state that the meaning of any social norm is that behaving according to it makes one conform to society. That also illustrates an important difference between conventions and social norms. Conventions are based on individual preferences and the convention emerges from these preferences in combination with the environment, which creates negative externalities for non-conforming behaviour. In contrast, the social norms are based on the fact that conforming to the majority in a society is seen as inherently good, even though there might not be a direct negative consequence for non-conforming behaviour for the individual.
 Moral norms do have some sort of extra meaning, their meaning is more restricted than for a social practice. In these cases, the meaning of the moral norms is simply promoting both the norm, and the associated values, and no other purposes are associated with following the norm. However, here the same caveat holds as for the social practices, in that the meaning can only be inferred from the behaviour that is shown, and not from its motivation, so other people will likely use their moral norms to assign meaning to the behaviour of others.

 A second difference between social practices on the one hand and conventions and norms on the other hand is that practices are very situated while conventions and norms can carry over many situations. Conventions mainly focus on the actions and behaviour in the interaction. Thus they abstract away from time and place and very often from the people and resources. This means that they can be used in more general ways. A similar argument holds for social norms that often just regulate a particular behaviour of a role. Thus any individual playing that role should comply to the norm. Complying to the norm will take care of smooth interactions, which are thus not explicitly described.

 A third difference between the three types of social rules lays in the social consequences (effects) of obeying the social rules. When an individual obeys a convention there is no direct positive social effect. The reason being that obeying a convention is seen as coming out of self interest. Given that everyone else obeys the convention, the best thing to do is to obey it as well. Thus there needs to be no social reward as the individual behaves out of pure self interest.\\
 This is different for obeying a social norm. If an individual obeys all the social norms she can be seen as a good member of society. I.e.\ obeying the social norm shows that you have the interest of society in mind (as well). This is in itself seen as positive and thus has a small positive effect.\\
 Obeying a social practice has a stronger effect. There is no obligation or self interest to follow the social practice. Thus, in principle, the choice to follow the social practice can be made out of free will. Thus it is a positive choice for this social practice. Therefore the reward of obeying the social practice is that the person gets attributed a status as promoting the meaning of that social practice. Thus, e.g.\ eating vegetarian could give one a status as promoting the environment and healthy living.

 A last difference that we discuss here is apparent when an individual breaks the social rule. If an individual breaks a convention while the other people all keep to it, the direct result of that behaviour for the individual will be negative. E.g.\ walking left while everyone else keeps to the right will make you bump against many people or having to walk slow and going around all people coming towards you. Thus the reaction on violation comes directly from the negative reward through the environment.\\
 When an individual violates a social norm such a direct negative consequence might not exist. E.g.\ if a person stops in the middle of the street to greet a friend and blocks the way for some time, he might not feel any direct negative consequence because people will move around them. However, people might also tell them to move to the side in order to let people move easy. So, violating a social norm will lead to sanctions from other people. This can be verbal or physical and direct or indirect, depending on the situation.\\
 Finally, when an individual does not conform to a social practice the reaction does not have to be negative at all. This does depend on the context in which the violation happens. Depending on the role and social status of the violator the other participants can try to adapt within the practice to solve the situation. If this does not work the participants can also switch to an alternative practice or even have an interaction that is not guided by any social practice. Thus violation basically means that an expectation is not fulfilled and all parties will take action to solve the situation. However, there is no special onus on the violator!

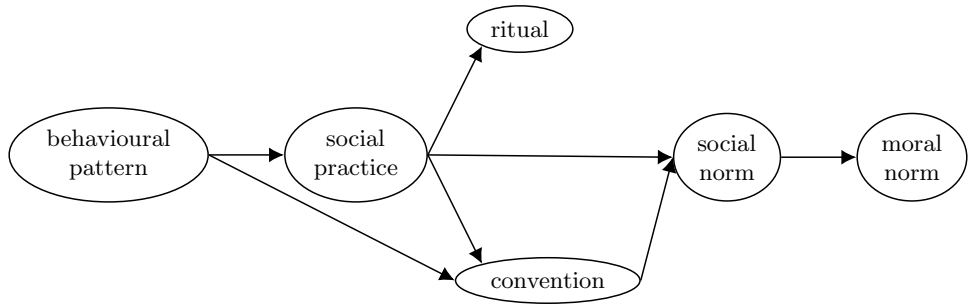
\begin{figure}[!ht]
     \centering
     \begin{tikzpicture}[line width=0.2mm, ball/.style={ellipse, draw, align=center}, >={Latex[length=2mm, width=2mm]}]
         \node[ball] (bp) {behavioural\\ pattern};
         \node[ball, right=of bp] (sp) {social\\ practice};
         \node[ball, below right=of sp] (c)  {convention};
         \node[ball, above right=of c]  (sn) {social\\norm};
         \node[ball, right=of sn] (mn) {moral\\norm};
         \node[ball, above right=of sp] (r) {ritual};
         \draw[->] (bp.east) -- (sp.west);
         \draw[->] (bp.east) -- (c.west);
         \draw[->] (sp.east) -- (c.north west);
         \draw[->] (sp.east) -- (sn.west);
         \draw[->] (sp.east) -- (r.south west);
         \draw[->] (c.east) -- (sn.west);
         \draw[->] (sn.east) -- (mn.west);
     \end{tikzpicture}
     \caption{Paths of emergence}
     \label{fig:emergence}
 \end{figure}

 In the final part of this section we will look at the connections between the different social rule types and especially how one type can evolve from another. In figure~\ref{fig:emergence} we offer up a tool for thinking about the emergence dependencies, which should not be seen as a strict characterization. As habits mainly emerge from repetition and automization and not specific social rules they are not incorporated in the figure. The figure should actually be read from right to left. An arrow between rule types means that the rule at the end of the arrow \emph{can} emerge from the rule at the start of the arrow. Thus (behaviour based) moral norms can only emerge from social norms, but conventions can emerge from social practices and directly from behavioural patterns. Rituals do not directly emerge from behavioural patterns. We do not claim that it is in no case possible, however we do claim that rituals needs a social environment to emerge from and that the step from behavioural pattern to ritual is generally too big, while from social practice to ritual is quite possible. Conventions (according to us) do not emerge from social norms, but social norms can emerge from conventions. The arrows do not mean that moral norms are the ultimate social rules that all social rules evolve into. The evolution can stop at any node!

 In the figure we have included the behavioural pattern, even though this is in itself not a social rule. We see a behavioural pattern just as any repeated behaviour, which therefore does not need social aspects. It is included in the figure since behavioural patterns can be the source of both conventions and social practices, in which it will take up the social aspects that it did not have before.
 The behavioural pattern is behaviour in which other persons are not treated as humans with their own goals and feelings, but rather as physical objects. For example when pushing away other persons to reach a certain place. The people are seen as obstacles and not as individuals who would also want to go to a specific place. The behavioural pattern is context dependent. The pushing behaviour only occurs in crowded places and disappears when there is enough space to pass other people. This contrasts with the social rules such as social practices which are performed together with other people and knowledge of their intentions is assumed. Conventions are also influenced by the goals or intentions of other people.\\
 The figure shows that rituals do not directly emerge from behavioural patterns. In general rituals need a long period of repetition of the joint behaviour and slow detachment of the actions from their original functional effects. It is quite impossible for this to happen if the context is determining the behaviour. It is easier to see rituals emerging from social practices where especially the performed actions within the practice become fixed and also fixed in order. In that sense rituals are more constrained than social practices and will emerge from social practices that have been performed for a long period between certain people.

 If we have to pick the most fundamental \emph{social} rule we would choose the social practice as it has the least constraints, largest set of possible behaviour and the other social rules can emerge from it.

 We have positioned conventions out of line with the other social rule types. This is to emphasize the fact that conventions only emerge in some type of contexts. In our example of a person meeting her friend in the street and stopping in the middle of the street to have a chat, there is no need for them to move to the side of the street. This would not be better for them given that other persons keep doing what they do (which is walking along while avoiding bumping into other people). However, it would be more convenient for the other people and thus might emerge as a social/moral norm. So, not \emph{all} social norms emerge through conventions.\\
 In general going from left to right in the figure there is an increase in frequency of the interactions that the rule governs and also the number of people that perform the actions. Behavioural patterns can be limited to one or two persons and take place very infrequent. E.g.\ whenever we have visitors from abroad we rent a car. This might happen a few times a year, but we do it consistently.
 Social practices require a higher frequency of the interaction. E.g.\ me and my wife always walk hand in hand. This is frequent, but only involves two persons. This behaviour can evolve into a convention and/or social norm when many people in the society (or at least the relevant part of it) are behaving like it. When the social rule becomes a social norm it also becomes an explicit social concern. The social norm can become a moral norm if the norm keeps stable over a longer period and detaches from the specific context and attaches itself to a value (or set of values).

 \subsection{Social practices and habits}
 Habits literature usually describes individual habits that are generally not very dependent on other people. Here lies a difference with the social practices. As descriptions of social practices actually include other people than the main actor and include their role, function and social meaning within the practice. In habit literature it is different, e.g. a person can have the habit of going to the supermarket to get his breakfast. He would have to interact with the cashier but the cashier's motivations and goals would not be evaluated. While in social practices such as a soccer match other actors have a role, so a soccer player will know that another actor such as a keeper will try to stop him from scoring by catching his shots on the goal.

 Complexity is different as well, it is only the less complex tasks which make it into habits. Therefore habits could not include complex human interaction i.e. seeing the other persons as their own actors with their own goals. For complex tasks it is not a habit that forms but there is goal-directed automaticity~\cite{lally2010}, however in this paper we will keep habit definition to only include simple tasks.

 How do habits and social practice overlap? In figure \ref{fig:combined_social_rules} they are both close to each other on the vertical axis, habits leaning a bit more to function and social practices a bit more to social meaning. Both social practices and habits are also very dependent on contextual cues. The cues define when a certain habit should be activated or when a practice can be applied. Because of this, habits can be seen as less complex social practices with the absense of other complex actors.

 It may happen that a habit becomes a social practice when other people start to join. For example a person that has a habit of taking a walk as a break during work. At some point colleagues may join the walk regularly which makes the event transcend from habit to social practice. A social practice becoming a habit is less likely, unless the activity still has value for the person engaging in the practice when others do not participate. Thus a person may still go for a walk when colleagues do not join. However a practice such as giving a handshake cannot turn into a habit as it needs another person.

 \subsection{Social practices and rituals}
 When a social practice is violated it leads to adaptation of the practice. For example doing a football match where there is only one goal instead of two goals. This is generally not a problem if the people performing the practice together agree. Violation of steps in a ritual does have a penalty, however, purely a social one. Examples of this range from simple sanctions, when you greet someone and the other person does not greet you back the other person could be considered rude, and could loose status. To more complex, e.g. in a church ceremony one who without reason violates the steps or is very noisy. This could lead to exclusion of the ceremony and even the group.
 Rituals regulate interactions of people~\cite{Mascarenhas2009}, they are based on their social meaning but could be in some cases individual. Social practices are both about the social and the functional. However as we have stated earlier, social practices do not just describe individual action.

 One could say that a social practice can become a ritual. For example in the Sint-Maarten begging party tradition. Which originated from the past, where the poor would go to the doors of farmers to beg for food to survive the winter. However in contemporary times 'most' people in the Netherlands do not have to beg for their food. Now the begging party consists of children going door to door for candy. In other words, the functionality mostly fell away, however people still do it because of the social meaning, e.g.\ honouring your traditions. Going from ritual to social practice may be possible when a ritual becomes used for a functional purpose. One example that comes to mind is a corrupted church ceremony where a preacher grows extremely rich from the donations. Where it traditionally was a ritual with a spiritual meaning (thus social meaning), the corrupted ceremony now became a way of making the preacher very rich which is functional for the rich preacher.

 \section{Implementing Social Rules}
 The connection and comparisons we have made in the previous section give us some implications for implementations. For example conventions are clearly connected to game theory~\cite{Lewis1969}. As we can see in figure~\ref{fig:combined_social_rules} rituals and conventions are very different on the sociality axis but on the same height on the individuality axis. Rituals have little or nothing to do with game theory. So the use of game theory seems mostly related to the fact that behaviour is primarily related to individual preferences and thus to individual gain or loss. Thus game theory will only be of benefit for implementing conventions while not very usable for the other social rules.

 Besides this feature of social rules we will also consider whether the frequency and repetitiveness of the rule is of big importance and whether the portion of individuals following the rule has influence on the adherence to the rule.

 Stating that game theory can be used to support implementation of conventions is easy, but it is a different matter to see \emph{how} this should be done. Let us take the example of walking on the right side of the street as use case. If all agents have a preference to walk on the right side of the street one will undoubtedly see the behaviour belonging to the convention. However, if I change the preference for all but one agent to walking on the left, the last agent will still keep walking on the right side. Its preference does not depend on the preferences of the others. So, the preference should be devised in a way that the convention follows from it, rather than just expressing the behaviour of the convention.\\
 This could be accomplished by something like having a preference to move in the same direction or walking as much as possible in straight lines while avoiding bumping into others. Notice, this could also lead to everyone walking on the left. This can be remedied by giving a bias that if everything else is equal one has a bias to walk on the right.

 Social practices are maybe the most difficult social rule when considering their implementation. A social practice mainly is a repeated behaviour affecting two or more parties. Thus their usefulness depends mainly on the frequency with which they are used. This entails a learning mechanism inside the agents that keeps track how successful and how useful the social practice was. I.e.\ did all parties fulfil the expected behaviour and did the use of the social practice lead to efficiency of deliberation and planning. Social practices can be used by only two persons or a small group. Thus they do not have to be widespread. When they do get widespread their social meaning gets more important and more abstract from the individual. In those cases it seems warranted to have the social meaning of the social practice implemented as a common structure available to all individuals of the application. If this is the case a social practice can be implemented as plan patterns available for the agents that have some fixed social meaning attached with them (thus saving deliberation over social and functional aspects separately).

 Depending on the applications, norms can be implemented in various ways. If all the agents are supposed to adhere to the norms and never break them, then they can be used as constraints on the behaviour and planning of the agents. However, if you expect to have rule breakers as part of the application, as might happen in a heterogeneous multi-agent system where you are not in control of all the agents, or if you want to build a social simulation, then you will have to take a slightly different approach. In that case the agents will have to incorporate the norms in their planning and you have to account for who sanctions the violators of norms. Are these other agents in the application, or some centralized system?

 Of these two options, a centralized system is easier to implement, since that would only require one system to detect norm violations and deal out sanctions. For legal norms this would work well, since that is how they work in human societies. However, for moral norms, which are more personal, or social norms, where sanctions tend to be social as well, this becomes harder to implement, meaning that agents will have to sanction the other agents themselves. This will require each agent to have a mechanism for detecting norm violations by other agents, and take action when they notice a violation. These two ideas can also be combined into one, by making the agents sanction each other, but include one or more agents whose only goal is to sanction others for breaking legal norms.

 In order to have norms affect an agent's planning, each agent has to know about the norm, which actions it allows and forbids, and the possible consequences of both following the norm and violating the norm. For legal norms this is easier to do, since they are normally already phrased in this manner, but for moral and social norms this is harder to achieve. Social norms in particular also involve some theory of mind on the part of the agents to see which norms are actually social norms and which agents care enough to sanction them. This also means that the outcomes of the actions have to be learned in some way, since they can change over time. If there is no norm evolution in the application, then this is not necessary since certain assumptions can be made.

 When a habit is activated a fixed sequence of actions is performed. For example going to the shop to buy breakfast contains: 1) go out, 2) walk to shop, 3) select food, 4) pay food (at cashier), 5) walk home, 6) go inside. Even though there is an external human actor in an implementation one does not have to consider complex interactions with that human, the person is merely there as one of the steps of the habits. This makes agents who can perform an individual plan such as BDI or 2APL sufficient for a habit implementation. They would only look at their own plan and would not consider the motivation and roles of others. Another approach to adding habits involves a learning curve, where actions (usually less complex actions) that people often do become more and more automatic the more often you do them~\cite{lally2010}. This can be used to incorporate habits that can emerge and change over time.

 Rituals are usually implemented more as a plan that agents follow step by step~\cite{Mascarenhas2009}. This seems to be comparable with habits with the only difference that there is added interaction in rituals. However this interaction can be modelled as cues which initiate the further progress of the plan, therefore the implementation of rituals will not be very different from habits. Context is important for rituals as this defines which ritual you are in and what plan to follow. If you are in an unknown context it is usually a good idea to use the heuristic of following what other do.

 In the previous section we already started the discussion on how the social rules can emerge from other social rules. We claimed that social practices can be seen as the most fundamental social rule. This means that other social rules can emerge out of social practices. Social practices have the least constraints while the other social rules have more constraints which decreases possible behaviour. From a sociological perspective by making more restrictions and adding frames we can start to understand human behaviour. However from an implementation perspective the opposite is actually going on. Before implementing no behaviour is possible, while by defining and implementing more possible behaviour the action space can be enriched. This makes social practices one of the most complex social rules to implement as it is the one that is the least precisely defined, where its execution can be context dependent and many aspects are implicit. This may feel counter intuitive as a primitive is usually the simplest element to implement or use in a programming language.

 While we argue that social practices is the fundamental social object where other social rules emerge from, this does not necessarily mean that it is practical to start implementation from the social practice, as it is not the simplest but rather the most complex. One possible way of implementation would be to go the other way around and start with the most constraining social rules such as norms and rituals and bringing them together for the social practices implementation.

 \section{Conclusion}
 We discussed a number of types of social rules and related them on a conceptual level. We have explicitly refrained from giving formalizations of the different social rule types, because the formalizations that exist already implicitly contain many decisions on which properties of the social rules are important. By having this discussion on a conceptual level it becomes possible to see which properties are important to distinguish the different types of social rules. However, we realize that the discussion also is somewhat abstract. Follow up steps need to be taken to translate the findings in the formalisms that already are used to model the social rules. This will give the discussion more rigour.

 We discussed some meta theorical considerations on how the social rules are related. Comparing their properties in terms of what is alike and what is different sharpens the initial definitions of the social rules. For example comparing social practices with habits forced us to think about whether a social practice is only social or could also be individual. After defining the similarities and differences of the social rules we have defined which techniques could be used if one wishes to implement social rules and move between them.

 We also gave some starting points on where to begin implementation of the social rules. Social practices could be seen as the most fundamental social rule however also the most complex to implement. Because the other social rules are more precisely defined and have less implicit and dynamic characteristics it follows that if we have a good framework to implement the social practices, the other social rules can be implemented by leaving some parts out, instantiating parameters and/or limiting some elements to fixed values.

 This work could be used as foundation for a methodology, to analyze the social rules, formalize the social rules and for an implementation of a framework. Figure \ref{fig:combined_social_rules} could be of help as it shows the general gist of this paper. What would be interesting to look into now is context, as it seems to be very important for habits, social practices and rituals to be able to see the context in which an agent is situated. This was out of the scope for this paper but would be necessary to do research in before implementing the here mentioned social rule types.

 We started the paper with the question whether all different social rule types can be derived from a fundamental social object. We did not explicitly answer this question in the paper. However, in figure 2 we indicated that there is a certain type of natural progression of social rules starting from behavioural patterns to legal norms. It indicates that at least some rules can progress through this network and pick up more restrictions and additional features along the way. E.g. keeping right in traffic can be a behavioural pattern that just happens to be true, it can grow into a convention if the expectation that other people keep to the right as well is added. It can become a moral norm if one could argue that keeping right is intrinsically a good thing to do (and is better than keeping to the left). And finally the same rule can become a legal norm by adding precise conditions on when and for whom it holds, what happens when it is violated etc.

\bibliographystyle{splncs04}
\bibliography{socialrules.bib}
\end{document}